\documentclass[preprint]{aastex631}
\usepackage{url}
\usepackage{chngcntr}

\accepted{June 11, 2021}

\submitjournal{ApJL}

\shorttitle{Magnetic Reflection at Collisionless Shocks}
\shortauthors{MADANIAN et al.}
\graphicspath{{./}}

\begin{document}

\title{Direct Evidence for Magnetic Reflection of Heavy Ions from High Mach Number Collisionless Shocks}

\correspondingauthor{Hadi Madanian}
\email{hmadanian@gmail.com}

\author[0000-0002-2234-5312]{Hadi Madanian}
\affiliation{Southwest Research Institute, 6220 Culebra Rd, San Antonio, TX 78238, USA}
\author[0000-0003-0682-2753]{Steven J. Schwartz}
\affiliation{Laboratory for Atmospheric and Space Physics, University of Colorado, Boulder, CO 80303, USA}
\author[0000-0003-4101-7901]{Stephen A. Fuselier}
\affiliation{Southwest Research Institute, 6220 Culebra Rd, San Antonio, TX 78238, USA}
\affiliation{University of Texas at San Antonio, San Antonio, TX 78249, USA}
\author[0000-0002-8175-9056]{David Burgess}
\affiliation{Queen Mary University of London, London, UK}
\author[0000-0002-2425-7818]{Drew L. Turner}
\affiliation{Johns Hopkins University Applied Physics Laboratory, Laurel, MD 20723, USA}
\author{Li-Jen Chen}
\affiliation{NASA Goddard Space Flight Center, Greenbelt, MD 20771, USA}
\author[0000-0002-7318-6008]{Mihir I. Desai}
\affiliation{Southwest Research Institute, 6220 Culebra Rd, San Antonio, TX 78238, USA}
\affiliation{University of Texas at San Antonio, San Antonio, TX 78249, USA}
\author{Michael J. Starkey}
\affiliation{Southwest Research Institute, 6220 Culebra Rd, San Antonio, TX 78238, USA}

\begin{abstract}

Strong shocks in collisionless plasmas, such as supernovae shocks and shocks driven by coronal mass ejections, are known to be a primary source of energetic particles. Due to their different mass per charge ratio, the interaction of heavy ions with the shock layer differs from that of protons, and injection of these ions into acceleration processes is a challenge. Here we show the first direct observational evidence of magnetic reflection of alpha particles from a high Mach number quasi-perpendicular shock using in-situ spacecraft measurements. The intense magnetic amplification at the shock front associated with nonstationarity modulates the trajectory of alpha particles, some of which travel back upstream as they gyrate in the enhanced magnetic field and experience further acceleration in the upstream region. Our results in particular highlight the important role of high magnetic amplification in seeding heavy ions into the energization processes at nonstationary reforming shocks.
\end{abstract}


\section{Introduction} \label{sec:intro} 

Fast moving collisionless shocks in space and astrophysical plasmas are regions of wave-particle interaction, energy transfer, and generation of highly energetic ions. Particle energization occurs through different acceleration processes, some of which become effective when seed ions obtain a minimum energy to engage in the process \citep{Marcowith2020Multi-scaleSystems,Caprioli2015SIMULATIONSSHOCKS}. The mechanisms by which ions from the bulk or thermal plasma are pre-accelerated to seed the acceleration process has been a point of debate \citep{marcowith_microphysics_2016}. Seeding and energization of heavy ions are of particular interest. Analysis of emission spectra from astrophysical shocks has shown mass-per-charge dependent preferential heating of heavy ions occurs in the post shock plasma \citep{Miceli2019Collisionless1987A}. In addition, abundance enhancement of heavy ions in galactic cosmic rays are believed to be caused by their acceleration at astrophysical shocks, and several modeling studies have been able to adapt mass dependent injection schemes to accelerate ions from a thermal ion pool \citep{Eichler1981AbundanceShocks,Caprioli2017ChemicalSimulations}. At high Mach number shocks, part of the upstream bulk flow energy is dissipated by the cross-shock electrostatic potential through ion reflection \citep{Hudson1965ReflectionShocks,Paschmann1982}. Ions with a kinetic energy less than this potential will be reflected back upstream, while directly transmitted ions are decelerated. Depending on their trajectory, reflected ions can be accelerated by the motional electric field ($ \textbf{E}=-\textbf{V} \times \textbf{B}$) or by the electric field within upstream perturbations. Reflected ions can also cause high amplitude magnetic enhancements in the upstream region that lead to nonstationarity and quasi-periodic reformation of the shock front \citep{hellinger_reformation_2002,Burgess2016MicrostructureShocks,Caprioli2014SIMULATIONSAMPLIFICATION,Bell2004TurbulentRays}, and reforming shocks in general exhibit a more intense magnetic amplification at the shock front  \citep{Russell1982OvershootsShocks,sulaiman_quasiperpendicular_2015,Madanian2021TheObservations}. Such strong fields can have an influence on the trajectory and the subsequent acceleration efficiency of upstream ions \citep{Giacalone1991EffectAcceleration,Caprioli2014SIMULATIONSDIFFUSION}.

The plasma of interplanetary shocks and planetary bow shocks is dominated by protons. Alpha particles constitute the most abundant minor heavy ion species with typical densities less than five percent of the proton density \citep{Kasper2007SolarCycle}. Due to their different mass per charge ratio, alpha particles do not experience the same deceleration by the cross-shock electrostatic potential as protons, which leads to a weaker deceleration of alpha particles relative to protons \citep{Gedalin2020PreferentialShocks}. Since the potential self-adjusts to process the main proton component, it is insufficient to reflect alpha particles. Directly transmitted alpha particles farther downstream of the shock form a ring beam distribution, which is unstable in the background drifting proton plasma and become isotropized into a shell-like distribution \citep{Fuselier1988AMPTE/CCEMagnetosheath}. Several studies on the other hand, have reported on the presence of nonthermal alpha particles upstream of Earth’s bow shock \citep{Ipavich1984CorrelationWind,Fuselier1990SpecularlyShocks,Fuselier1995SuprathermalRegion,broll_mms_2018}. It was initially thought that ion leakage from the heated plasma in the downstream region produces these ions. However, later investigations which compared He\textsuperscript{2+}/H\textsuperscript{+} density ratios in the energized and thermal ion populations determined that these ions originate from the solar wind.

Up to date, there have been no direct measurements to bridge the knowledge gap on how heavy ions reflect at high Mach number shocks. In what follows we show direct observational evidence for magnetic reflection of alpha particles at high Mach number quasi-perpendicular shocks using high resolution in-situ spacecraft observations.
\section{Data Sets} \label{sec:dataset}
We use in-situ measurements from the Magnetospheric MultiScale (MMS) spacecraft \citep{Burch2016MagnetosphericObjectives}. Ultra high time resolution of MMS observations at Earth’s bow shock have enabled studying new interesting microphysical shock processes. In our analysis, magnetic field vectors are from the fluxgate magnetometer \citep{Russell2016TheMagnetometers}, electric field data from the electric double probe (EDP) experiment \citep{Ergun2016TheMission,Lindqvist2016TheMMS}, and 3D ion measurements from two separate instruments: the fast plasma investigation (FPI) instrument \citep{pollock_fast_2016}, and the hot plasma composition analyzer (HPCA) \citep{Young2016HotMission}. The FPI ion measurement cycle is 150 milliseconds with no mass-per-charge differentiation. The HPCA obtains distributions of five major ion species every 10 seconds. We use HPCA data to confirm the presence of reflected alpha particles, while FPI data provide the necessary tool to investigate detailed microphysics of the reflection processes. Although the FPI does not distinguish between ion species, signatures of cold ion populations for both protons and alpha particles can be distinguished in the distribution. Since protons and alpha particles move together in the solar wind flow, they are expected to have a similar trace and trajectory in the velocity space. However, due to their different mass per charge ratios, their associated beams appear at two separate energies in the FPI electrostatic analyzer, and at separate velocities in the velocity space. The FPI also measures the electron pressure tensor which we use to obtain the ambipolar electric field. Results in the paper are based on satellite 1 (MMS1) measurements, except spatial derivatives for which all four spacecraft data are utilized.

\section{Observations and Results}\label{sec:results}
\begin{figure}[ht!]
\plotone{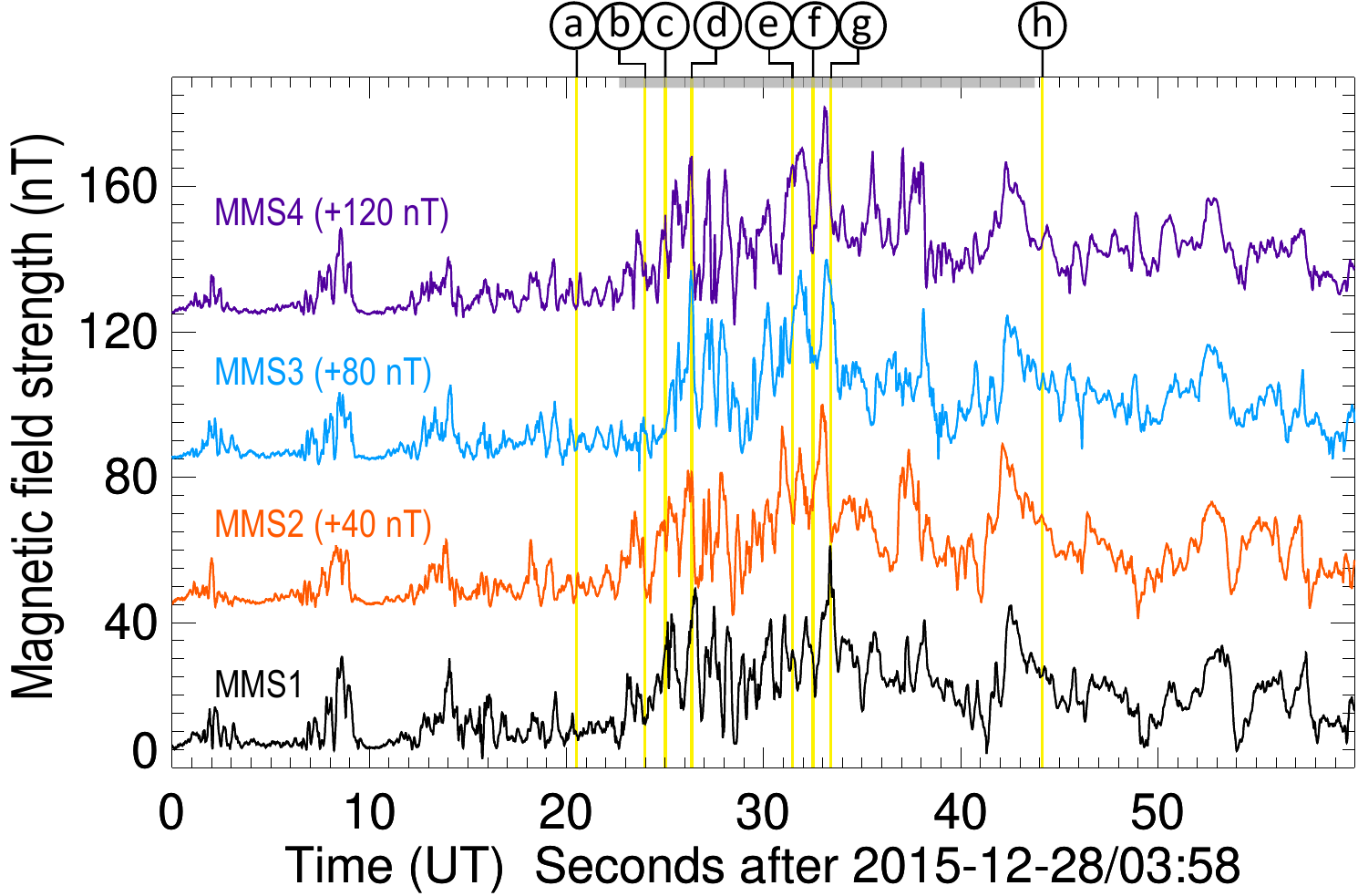}
\caption{Nonstationarity and magnetic amplification. Magnetic field profiles measured by four MMS spacecraft between 03:58:00 and 03:59:00 UT on 28 December 2015. The spacecraft have a close tetrahedron formation traveling near the orbit apogee. Data from MMS2 (orange), MMS3 (blue), and MMS4 (purple) are shifted in magnitude by 40, 80, and 120 nT respectively, to avoid clutter. The grey bar on the top marks the main shock transition period. Several timestamps are highlighted on this figure for further analysis of underlying ion distributions.
\label{fig:fig1}}
\end{figure}

Figure~\ref{fig:fig1} shows four MMS spacecraft observations of the magnetic field during a transition from the upstream solar wind to the downstream magnetosheath. The main shock transition layer, corresponding to the period of high magnetic fluctuations in data, is marked with a grey bar. Although the intra-spacecraft separation between different MMS pairs is on average only 25 km, each spacecraft sees a different shock profile, indicative of the spatial and time-varying nature of the shock layer. The magnetic field far upstream in the unperturbed solar wind is about 2.7 nT, while the highest magnetic field strength at the shock is 60.5 nT. The upstream plasma density is 10.5 $\rm cm^{-3}$ while the alpha particle to proton density ratio is 0.04. The shock has a quasi-perpendicular orientation ($\theta_{Bn}=83^{\circ}$) and travels in the upstream solar wind flow with an Alfvénic Mach number of $M_{Alf}=27$. While such a Mach number is relatively high for planetary bow shocks, it is considered fairly low in astrophysical shock studies. An earlier analysis of this shock event described its nonstationarity and reforming nature \citep{Madanian2021TheObservations}. It was shown that reflected protons generate magnetic enhancements in the upstream region that are convected towards the shock by the solar wind. They cause significant variation and intense magnetic amplification in the shock layer.

\begin{figure}[hb!]
\plotone{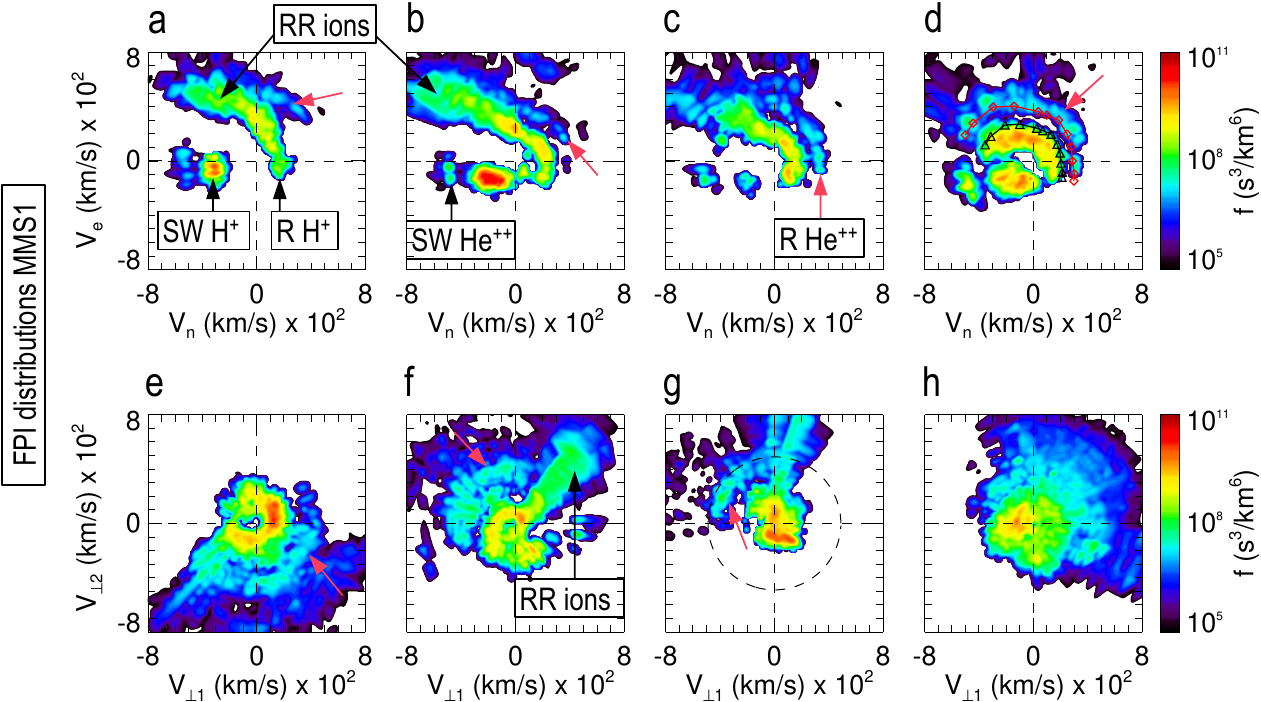}
\caption{Evidence of alpha particle reflection. This figure shows ion phase space density (s\textsuperscript{3}/km\textsuperscript{6}) distributions across the shock. Each distribution shows FPI data over a single measurement cycle. The corresponding timestamps are indicated in Figure~\ref{fig:fig1} with lower case letters. Distributions in panels (a - d) are shown in the NIF in a plane containing $\hat{n}$ and the upstream motional electric field $\hat{e}$. The third coordinate axis in this system is into the page, tangent to the shock surface, and along the projection of the upstream magnetic field. Distributions in panels (e - h) are shown in the local plasma frame in a plane perpendicular to the magnetic field. Ion populations of protons (H\textsuperscript{+}) and alpha particles (He\textsuperscript{2+}) are identified by SW (solar wind), R (reflected), and RR (returning reflected) labels. Red arrows indicate populations of non-gyrotropic alpha particles.
\label{fig:fig2}}
\end{figure}

Figure~\ref{fig:fig2} shows ion velocity distributions at several instances across the shock as measured by the FPI. The top row distributions are in the normal incidence frame (NIF) with the abscissa along $\hat{n}$ and the ordinate along the upstream motional electric field $\hat{e}$. Figure~\ref{fig:fig2}.(a) shows an ion distribution upstream of the shock, where the solar wind proton beam is labeled at $V_n\sim-350$ $\rm km~s^{-1}$. The arc-shaped feature in the distribution that begins at $V_n\sim200$ $\rm km~s^{-1}$ represents shock reflected solar wind protons and includes freshly reflected ions traveling sun-ward $(V_n > 0, V_e \sim 0)$, ions being turned around by the motional electric field $(V_n \sim 0, V_e > 0)$, and reflected ions that are further accelerated and are returning to the shock with $-V_n$ velocity component (RR ions). The distribution in Figure~\ref{fig:fig2}.(b) is near the upstream edge of the shock transition layer. The solar wind alpha particle beam is identified at $V_n\sim-500$ $\rm km~s^{-1}$. An interesting ion feature in Figure~\ref{fig:fig2}.(c) is a population of reflected ions at higher velocities than protons, which we associate it with alpha particles. Small traces of reflected alpha particles are also observed in Figure~\ref{fig:fig2}.(a,b) (marked with red arrows). The flux intensity of these ions progressively increases into the shock layer. In Figure~\ref{fig:fig2}.(d) we show that adjusting the mass-per-charge ratio of an ion at those velocities for an alpha particle results in a velocity trace similar to that of protons. Higher rigidity of alpha particles may allow them to experience a larger potential difference along the motional electric field. We select a number of points, where the two arcs in the velocity space are well distinguishable, marked by red diamonds. We transform these velocities to the energy space through the proton NIF, and remap the resultant energies back to the velocity space with the mass per charge ratio of an alpha particle. The results, marked with black triangles, falls along the same path as protons, though at slightly higher speeds. This agreement supports our claim that ions observed at higher speeds are indeed alpha particles that are processed as protons by the FPI. It should be noted that the NIF transformation also affects alpha particles differently than protons, which can contribute to the different speed of the alpha particle trace. These observations also agree with the HPCA measurements of He\textsuperscript{2+} ions acquired over a longer time period (see Figure~\ref{fig:figSM1} in the Appendix), confirming our interpretation of these features. 

Comparison of reflected ion populations in Figure~\ref{fig:fig2}.(a - d) reveals interesting differences between reflection of protons versus alpha particles. Nearly specular reflection of protons by the cross-shock electrostatic potential creates an ion signature with reversed $V_n$ velocity of the incident solar wind protons, as shown in Figure~\ref{fig:fig2}.(a – c), which are associated with newly reflected ions. Reflected alpha particles in Figure~\ref{fig:fig2}.(a,b) lack such an ion population and seem to be further along in their cycloidal motion. The ion populations in Figure~\ref{fig:fig2}.(d) are also rather distinct compared to previous upstream distributions in that, the energization of returning ions along the upstream motional electric field is lessened or diminished. Alpha particles exhibit non-gyrotropic distributions and appear in gyration in the enhanced magnetic field environment. Depending on their gyrophase and the dynamically variable background magnetic field topology, they can appear upstream of the shock. These observations are a direct indication that the trajectory of alpha particles and their reflection from the shock are driven by magnetic forces at the shock front. The dominance of gyrokinetic effects is supported by the intense magnetic amplification in this region. The corresponding magnetic field strength for distributions in Figure~\ref{fig:fig2}.(a - d) is 9.8, 12.4, 29.2, and 36.9 nT, respectively corresponding to a magnetic enhancement ratio of 3.7, 4.6, 11, and 13.9 from the upstream condition. However, it should also be noted that Figure~\ref{fig:fig1} shows that at timestamp (d), MMS3, which is 20 km behind MMS1 along the shock normal, is observing a significantly larger magnetic field of $\sim 57$ nT.

Gyrating alpha particles that are not reflected eventually propagate downstream. Figures~\ref{fig:fig2}.(e - g) show ion distributions deeper in the shock layer in a plane perpendicular to the local magnetic field. Noteworthy in these distributions is the non-gyrotropic alpha particle population denoted by red arrows, consistently observed at $|V|\sim400$ $\rm km~s^{-1}$. Figure~\ref{fig:fig2}.(e) also shows a population of dispersed energetic ions in the lower half of the plane. A possible source for these ions could be returning reflected ions that have been scattered by the shock or upstream structures. In Figures~\ref{fig:fig2}.(e) and (f), the intense ion fluxes around the origin occupying almost all four quadrants of the velocity space are directly transmitted protons which seem to be driven locally at the shock front (i.e., no acceleration comparable to the upstream motional electric field). The streak of ions in the $(+V_{\perp1},+V_{\perp2})$ quadrant in Figure~\ref{fig:fig2}.(f) also shows the returning reflected ions accelerated to different energies and in propagation towards downstream. The gyrophase range of alpha particles between Figure~\ref{fig:fig2}.(e) and (f) changes due to a rotation in the local magnetic field direction. We also note that the magnetic field at times (c), (d), (f), and (g) timestamps also has a substantial component perpendicular to the coplanarity plane. The distribution in Figure~\ref{fig:fig2}.(g) is measured at the highest magnetic field strength across the shock, as measured by MMS1. It shows an almost isotropic population of protons, a non-gyrotropic population of alpha particles, and returning reflected ions. The distribution in Figure~\ref{fig:fig2}.(h) is measured near the downstream edge of the shock transition layer showing an example of the fully shocked solar wind plasma where ions exhibit a filled-shell distribution.

\begin{figure}[ht!]
\plotone{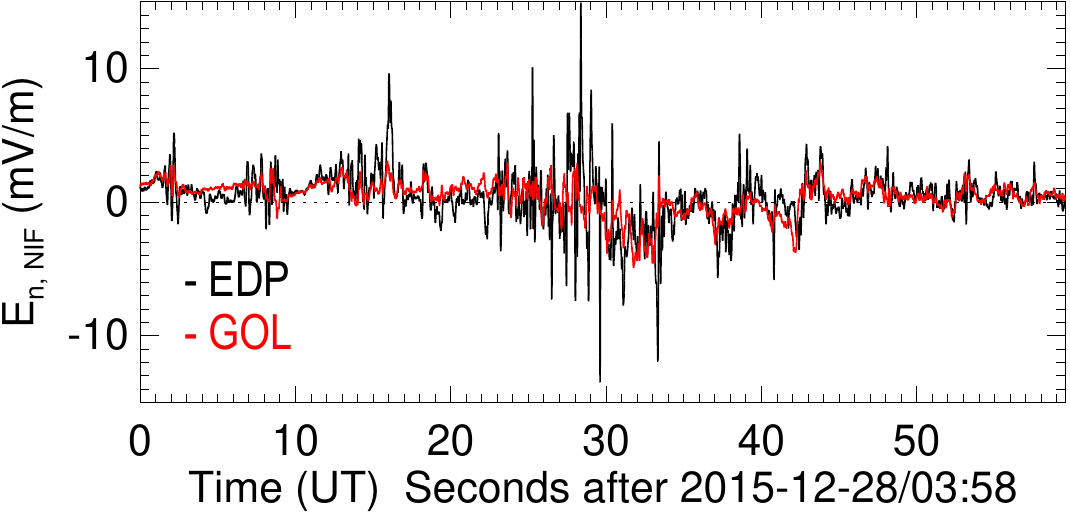}
\caption{Electric field variations across the shock. The electric field component along the shock normal vector obtained from two sources, 1): from measured by the EDP instrument onboard MMS1 (black), and 2) frpm the direct calculation of the Generalized Ohm’s Law using four spacecraft measurements (red). The two estimates overall are in good agreement except when spikey features in EDP data are present.
\label{fig:fig3}}
\end{figure}

The consistent presence of gyrating alpha particles with relatively constant gyration speeds across the shock, as shown in Figure~\ref{fig:fig2}.(e - g), indicates that heavy ions are less sensitive to the cross-shock potential than protons \citep{Gedalin2020PreferentialShocks, Fuselier1994HShock}. Alpha particles are also less likely to be scattered by waves at the shock which are excited predominantly at proton kinetic scales. In fact analysis of the electric field across the shock between 03:58:20-40 UT reveals that the normal component of the field changes constantly without an obvious nonzero trend, indicative of a continuous buildup of a potential (Figure~\ref{fig:fig3}). It is unclear how a cross-shock electrostatic potential varies at a nonstationary shock, since incident upstream perturbations can maintain a variable fraction of that potential. Therefore, it is difficult to infer a potential for these shocks through conventional methods. Nonetheless, the rate of change in the proton kinetic energy can be used as a proxy to estimate an “effective” cross-shock potential, and to determine its effect on the alpha particle beam (see Appendix \ref{sec:app2}). The upstream proton flow for this event has a speed of $V_{up, NIF}=474.1$ $\rm km~s^{-1}$ in the NIF, and observations downstream of the shock indicate of an asymptotic flow speed of $V_{dn}=108.7$ $\rm km~s^{-1}$ along the shock normal. The associated shock potential would decelerate alpha particles to $V_{\alpha}=343.4$ $\rm km~s^{-1}$, which is mostly the gyration speed. This speed in the FPI frame is marked with the dashed circle in Figure~\ref{fig:fig2}.(g), where our estimate from this simple approach seems to agree reasonably well with the gyration speed of the observed nongyrotropic alpha particles.

\section{Discussion and Conclusions}\label{sec:disc}
By using high time resolution in-situ spacecraft measurements we investigate the physical mechanism behind reflection of alpha particles from high Mach number quasi-perpendicular shocks. Detailed analysis of a case study shows that the shock is highly nonstationary and is dynamically modified by upstream transient perturbations. The highest magnetic amplification $(R_{Max})$ at the shock reaches more than 22 times the upstream magnetic field strength. In Figure~\ref{fig:fig1} we show that such an intense magnetic amplification is accompanied by upstream magnetic perturbation which are generated by ion kinetic processes and are further amplified at the shock \citep{Madanian2021TheObservations,sundberg_dynamics_2017,Caprioli2014SIMULATIONSAMPLIFICATION, sulaiman_quasiperpendicular_2015}. Our detailed analysis of ion distributions in Figure~\ref{fig:fig2}, upstream, within, and immediately downstream of the shock indicates that alpha particles create vortices in the velocity space as they gyrate in the enhanced magnetic field at the shock front. Depending on their gyrophase and the background magnetic field topology, some gyrating alpha particles travel upstream of the shock as reflected ions.

Extreme magnetic amplification at the shock front, much stronger than both upstream and downstream magnetic fields, is essential for the magnetic reflection process. In the stationary regime, the shock transition scale $(L)$ is on the order of the upstream proton convective gyroradius in the downstream magnetic field \citep{Bale2003Density-TransitionShocks}. The shock therefore appears as a discontinuity to heavy ions \citep{Gedalin2020Large-scaleMotion}. However, for the nonstationary shock considered here, the convective gyroradius of decelerated alpha particles in the intense magnetic field of the shock front $(\rho_{\alpha})$ is much smaller than the shock transition scale (Table~\ref{table:1}), which allows for effective gyromotion of alpha particles in that region. We also identify this behavior in two other shock events listed in Table~\ref{table:1}. For these shocks we observe: 1) shock front nonstationarity and high magnetic amplification rates, 2) reflected alpha particles upstream of the shock in HPCA data, and 3) signatures of non-gyrotropic alpha particle distributions at the shock front in FPI data. The relation between different scale parameters as discussed above is valid also for these shocks, which support our interpretation of the magnetic reflection process through these observations. The value inside the bracket in the last column of Table \ref{table:1} is the magnetic field used to calculate $\rho_{\alpha}$, taken at the center of the MMS tetrahedron and averaged over the compression region observed close to the shock front.

\begin{splitdeluxetable*}{lccccccBccccc}
\tablecaption{Plasma, shock, and scale parameters for three events
\label{table:1}}
\tablewidth{0pt}
\tablehead{
\colhead{Date/Time} & \colhead{$M_{Alf}$} & \colhead{$M_{fm}$} & \colhead{$\theta_{Bn}$} & \colhead{$B_{up}$} & \colhead{$R_{Max.}$} & \colhead{$B_{d}$} & \colhead{$V_{up, NIF}$} & \colhead{$V_{dn}$} & \colhead{$V_{\alpha}$}  & \colhead{$L$} & \colhead{$\rho_{\alpha}$} \\
\colhead{} & \colhead{} & \colhead{} & \colhead{(Deg.)} & \colhead{(nT)} & \colhead{} & \colhead{(nT)} & \colhead{($\rm km~s^{-1}$)} & \colhead{($\rm km~s^{-1}$)} & \colhead{($\rm km~s^{-1}$)} & \colhead{(km)} & \colhead{(km) [nT]}}
\startdata
2015Dec28/03:58:24 & 27.0 & 15.0 & 83 & 2.7 & 22.4 & 15.2 & 474.1 & 108.7 & 343.4 & 325.6 & 229.0 [31.3] \\
2015Dec28/04:32:04 & 20.1 & 12.8 & 68 & 3.0 & 19.2 & 12.6 & 472.8 & 99.6 & 341.6 & 391.7 & 233.1 [30.6] \\
2018Dec16/01:23:32 & 24.0 & 4.9 & 76 & 0.7 & 20.1 & 3.7 & 223.1 & 54.8 & 162.8 & 629.4 & 399.9 [8.5] \\
\enddata
\end{splitdeluxetable*}

The shock events in this study have a quasi-perpendicular orientation. However, at quasi-parallel shocks steepening of upstream ultra-low frequency waves also creates large amplitude magnetic pulsations and intense magnetic amplification at the shock front \citep{Burgess2013MicrophysicsPlasmas, Liu2021MagnetosphericWaves, schwartz_quasi-parallel_1991}, which suggests that magnetic reflection can also function at quasi-parallel shocks. Interestingly, several previous studies that report on the presence of reflected alpha particles upstream of the terrestrial bow shock have also been conducted at shocks that show noticeably high magnetic amplification and nonstationary features \citep{Fuselier1990SpecularlyShocks,broll_mms_2018}, although these effects were unrecognized or overlooked in those studies.

We report on the first observational study that provides rationale for reflection of heavier-than-proton ions at collisionless shocks in proton-dominated plasmas. Is is argued that, unlike protons, alpha particles are reflected through gyration in the dynamically intensified magnetic field at the shock front. Since reflected ions seed the diffused ion population, our findings suggest that shocks that exhibit high magnetic amplification (i.e., reforming shocks at high Mach numbers,) are more likely to produce a diverse admixture of heavy ion species and protons in the diffused energetic ion population. These observations have important implications for theoretical and simulation studies of heavy ion acceleration at high Mach number astrophysical shocks \citep{Caprioli2017ChemicalSimulations,Caprioli2011Non-linearShocks,Meyer1997GalacticRatio}, generation of heavy solar energetic particles \citep{Yang2011AccelerationNonstationarity}, and heavy ion interactions with plasma structures inside the solar corona \citep{Zimbardo2011HeavyCorona}.

\begin{acknowledgments}
This work was supported in part by National Aeronautics and Space Administration (NASA) grants NNG04EB99C and 80NSSC18K1366. All data used in this study are hosted by NASA and are publicly accessible at (\url{https://spdf.gsfc.nasa.gov/pub/data/mms}). We thank the entire MMS team for providing the mission data.
\end{acknowledgments}

\appendix
\counterwithin{figure}{section}
\section{Electric field across the shock} \label{sec:app1}

The electric field (shown in Figure~\ref{fig:fig3}) can be calculated from the electron equation of motion and the Generalized Ohm’s Law (GOL):

\begin{equation} \label{eq:gol}
\textbf{E}_{GOL} = - \textbf{v}_{e}\times\textbf{B} + \frac{1}{en_e}\nabla . \textbf{P}_e + \textbf{V}_{NIF}\times\textbf{B}
\end{equation}

where $\textbf{v}_e$, $n_e$, and $\textbf{P}_e$ are the electron velocity, density, and pressure tensor respectively, and $e$ is the unit charge. The last term in Equation \ref{eq:gol} is the transformation field into the NIF. $\textbf{V}_{NIF}$ is defined by $\textbf{V}_{NIF}= \hat{n} \times (\textbf{V}_{SW,shock} \times \hat{n})$, where $\textbf{V}_{SW,shock}$ is the upstream solar wind velocity in the shock rest frame. We directly estimate the right-hand side of Equation \ref{eq:gol} using measured quantities. $\textbf{v}_e$, $n_e$, and $\textbf{B}$ are interpolated to the center of the MMS tetrahedron. The divergence term in Equation \ref{eq:gol} is the ambipolar field and is calculated using electron pressure tensors and spatial derivative techniques for multipoint measurements \citep{Chanteur1998SpatialAnalysis}. The off-diagonal terms in the electron pressure tensors are assumed to be small and negligible. Inertial terms in the GOL are also ignored. 

The electric field is also measured directly onboard by the EDP instrument. A low band-pass filter has been applied to electric field data to remove variations higher than 12 Hz, matching the electron sampling rate, before calculating the normal component. Ideally, the cross-shock electrostatic potential can be obtained by integrating the normal component of the electric field $E_n$ in the de Hoffman-Teller (dHT) frame \citep{DeHoffmann1950Magneto-HydrodynamicShocks}, in which the upstream solar wind flow velocity becomes parallel to the upstream magnetic field. However, for highly oblique quasi-perpendicular shocks, the electric field in the dHT frame is completely dominated by the transformation field and the shock potential cannot be determined through conventional methods \citep{Schwartz2021EvaluatingShocks}. In addition, the speed of the shock front at nonstationary reforming shocks is modualted by upstream cyclic perturbations, further complicating the spatial integration of the electric field.


\section{Ion Motion in the Downstream} \label{sec:app2}

We adopt a simple approach to determine the dominant velocity component of heavy ions at the shock front. The rate of change in the proton kinetic energy along the shock normal provides an estimate of the effective cross-shock electrostatic potential \citep{Lee2000HeatingCorona, Gedalin2020PreferentialShocks}, which can be obtained from:

\begin{equation} \label{eq:delphixshockpot}
e\Delta\Phi = \frac{1}{2} m_p (V^2_{up} - V^2_{dn})
\end{equation}

\noindent where $m_p$ is the proton mass, and $V_{up}$ and $V_{dn}$ are the upstream and downstream proton flow speeds along the shock normal, respectively. It is assumed that velocity variations due to $\Delta\Phi$ for all ions is along $\hat{n}$ (i.e., $V_{up}$ tan($\theta_{Bn}$) = $V_{dn}$ tan($\theta_{dn}$), where $\theta_{dn}$ is the downstream obliquity angle). We define $\epsilon$ as the ratio of the cross-shock electrostatic potential to the upstream proton kinetic energy. From~\ref{eq:delphixshockpot}, the downstream velocity of incident ion $i$ with mass M and charge state Q is defined as: 

\begin{equation}
V_{dn,i} = V_{up} (1 - \epsilon Q/M)^{1/2} = \alpha V_{up}
\end{equation}

In the dHT frame, the ion gyration speed $(V_{dn,g})$ and the variation in the field-aliened velocity component $(\Delta V_{dn, \parallel})$ immediately downstream of the shock are determined by the downstream velocity difference between the ion and the proton flow ($\Delta V_i$) through:

\begin{equation} \label{eq:deltaV}
\Delta V_i = V_{dn,i} - V_{dn} = V_{up}(\alpha  - \frac{B_{up,tang.}}{B_{dn,tang.}})
\end{equation}

\begin{equation} \label{eq:Vgyro_para}
(\Delta V_{dn, \parallel}, V_{dn,g}) = \Delta V_i (\text{ cos}(\theta_{dn}), \text{ sin}(\theta_{dn}))
\end{equation}

\noindent In these equations, $tang.$ and $n$ subscripts refer to the total tangential and normal components of the magnetic field. Equation \ref{eq:Vgyro_para} indicates that immediately downstream of high Mach number quasi-perpendicular shocks where $\theta_{dn} \rightarrow 90^{\circ}$, the ion motion is dominated by gyration. For shocks with high magnetic amplification, Equation~\ref{eq:Vgyro_para} may be rewritten as:

\begin{equation} \label{eq:expanded}
(\Delta V_{dn, \parallel}, V_{dn,g}) = [V_{up}(\alpha  - \frac{B_{up,tang.}}{B_{dn,tang.}})] (|\frac{B_{up,n}}{B_{dn,tang.}}|, 1)
\end{equation}

\begin{figure}[h!]
\centering
\includegraphics[width=0.8\columnwidth]{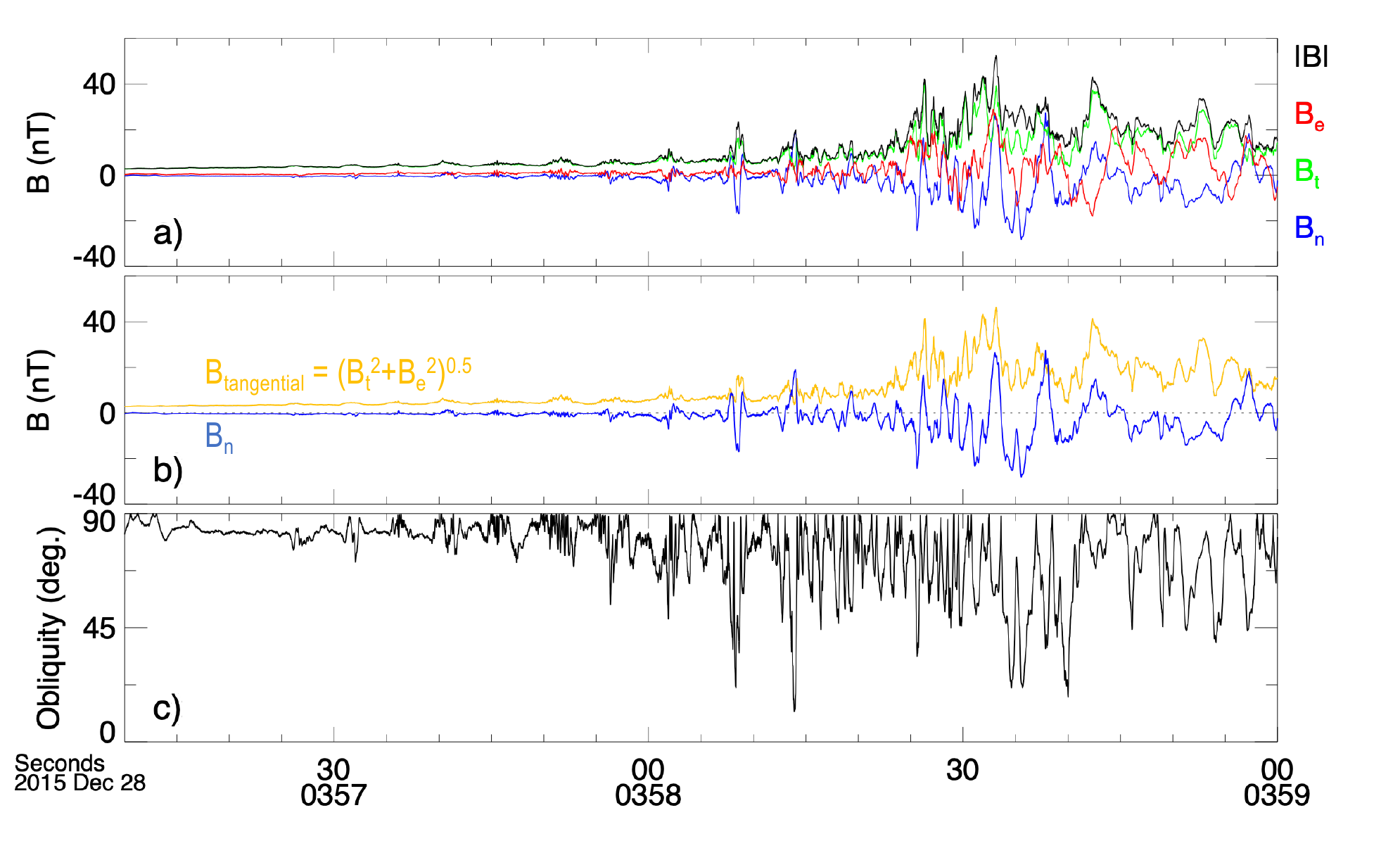}
\caption{Supporting figure for magnetic field components and local obliquity angle during the shock event on 28 December 2015 at 03:58:24 UT. (a) Magnetic field components interpolated to the center of the 4 spacecraft tetrahedron and shown in the shock normal coordinates, (b) the total tangential (yellow) and  normal (blue) components of the field, (c) the local obliquity angle.
\label{fig:b_component_timeseries}}
\end{figure}

\noindent It is worth noting that magnetic field ratios in Equation~\ref{eq:expanded} can be modulated by self-generated magnetic fields and nonstationarity effects at the shock. Panels (a) and (b) in Figure~\ref{fig:b_component_timeseries} show the components of the magnetic field tangent and normal to the shock surface for the event discussed in the manuscript.

\clearpage

\section{Supporting Figures}

\begin{figure}[ht!]
\centering
\includegraphics[width=0.8\columnwidth]{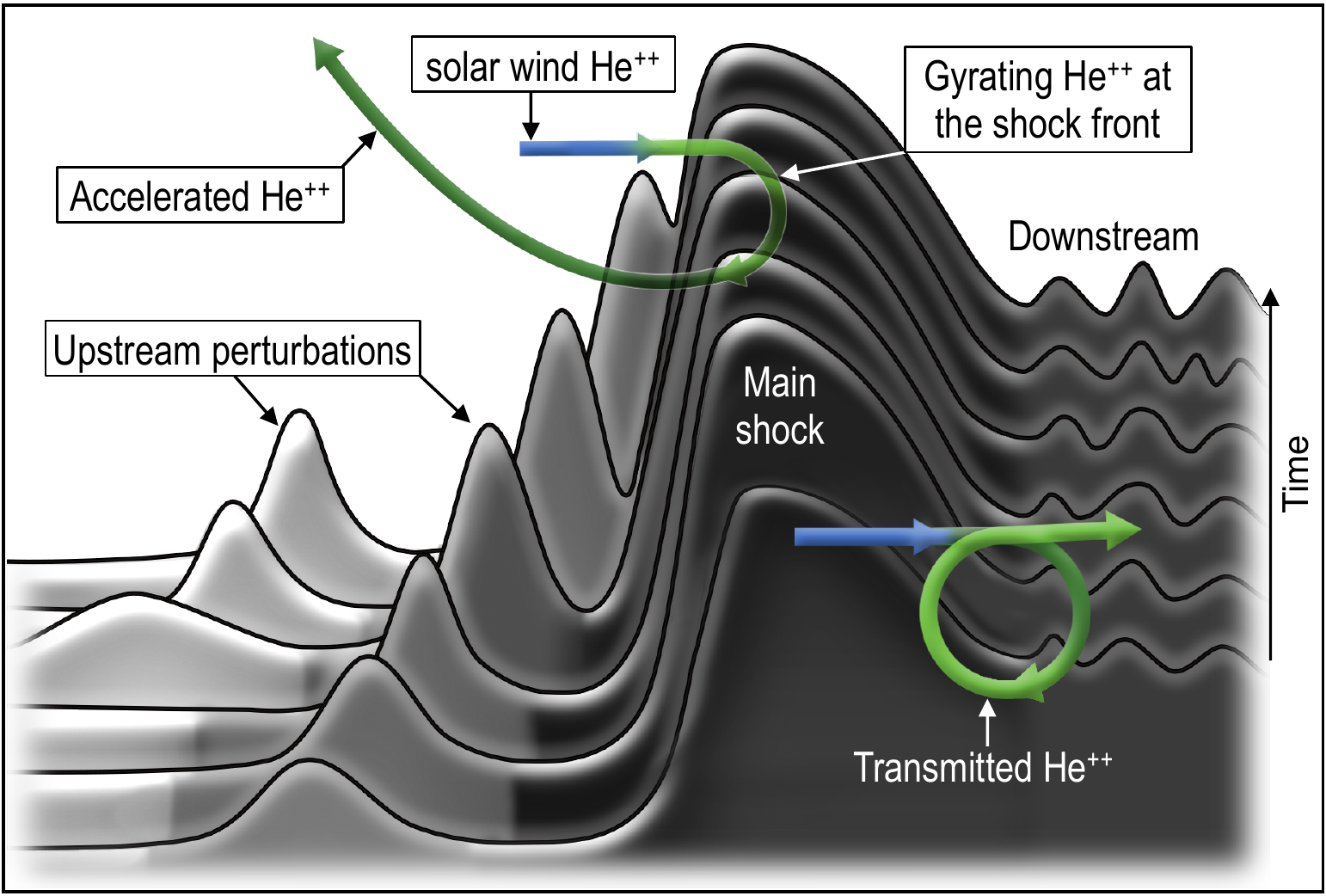}
\caption{A schematic illustration representing the magnetic field structure and alpha particle dynamics across the shock. The stacked magnetic field profiles (black lines) represent variations in the field strength across the shock front. The amplitude of each peak indicates the relative magnetic field strength at different times and locations. The actual magnetic field vector also changes in direction. Two example trajectories of reflected (top) and transmitted (bottom) solar wind alpha particles are also over-plotted to illustrate the ion behaviors. It should be noted that the ion dynamics have a different time dependence than the illustrated magnetic field variations.
\label{fig:figSM4}}
\end{figure}

\begin{figure}[h!]
\centering
\includegraphics[width=0.6\columnwidth]{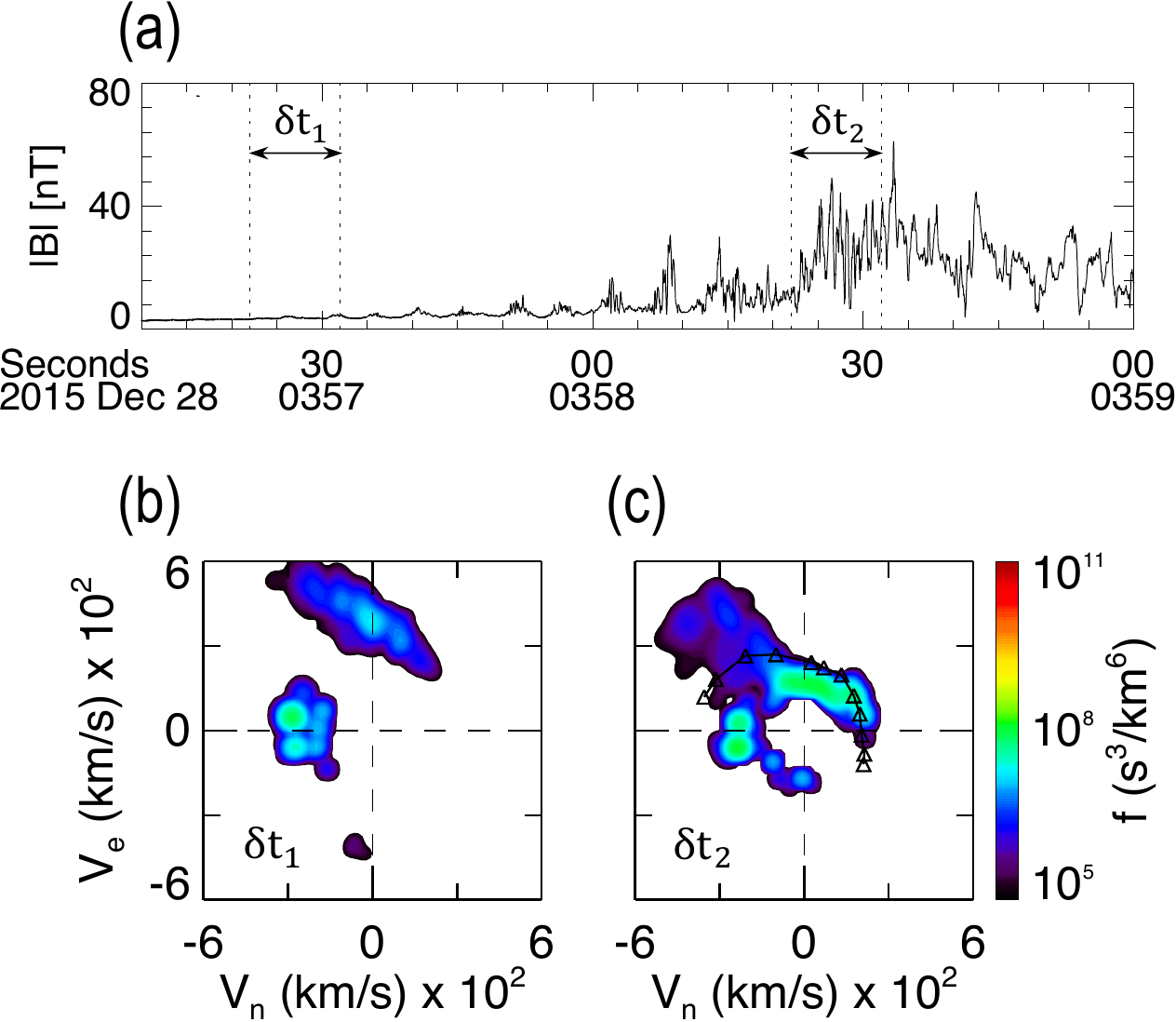}
\caption{ Supporting figure for the shock event on 28 December 2015, 03:58:24 UT. (a) The magnetic field profile across the shock. (b) He\textsuperscript{2+} distribution in the NIF measured upstream of the shock measured by the HPCA over one measurement cycle $(\delta t_1)$. In this panel, in addition to the solar wind alpha particle beam, a less intense population in the shape of an arc is observed with velocity component mostly along $+V_e$. These are reflected alpha particles that have been accelerated by the upstream motional electric field and are near the turning point of their gyration. (c) He\textsuperscript{2+} distribution measured across the main shock layer $(\delta t_2)$. In panel (c), a population of reflected ions with $+V_n$ velocity component propagating away from the shock is also visible. Ion energization along $+V_e$ is not as pronounced as the upstream distribution in panel (b). The black triangle trace, as discussed in Figure~\ref{fig:fig2}.(d), is overplotted on this panel for comparison. There exists a more energetic and dispersed ion population (the dark blue patch in the $(-V_n,+V_e)$ quadrant) which correspond to returning reflected ions. Ions with the $-V_e$ velocity component can be interpreted as gyrating particles. They can also be due to proton bleedover at low energies \citep{Young2016HotMission,starkey_acceleration_2019}. HPCA observations confirm the presence of reflected alpha particle upstream of the shock. However, the relatively long measuring cycle of HPCA (10 s) conceals finer details about the reflection mechanism, electric and magnetic field conditions, and subpopulations of ions. 
\label{fig:figSM1}}
\end{figure}

\begin{figure}[h!]
\centering
\includegraphics[width=0.95\columnwidth]{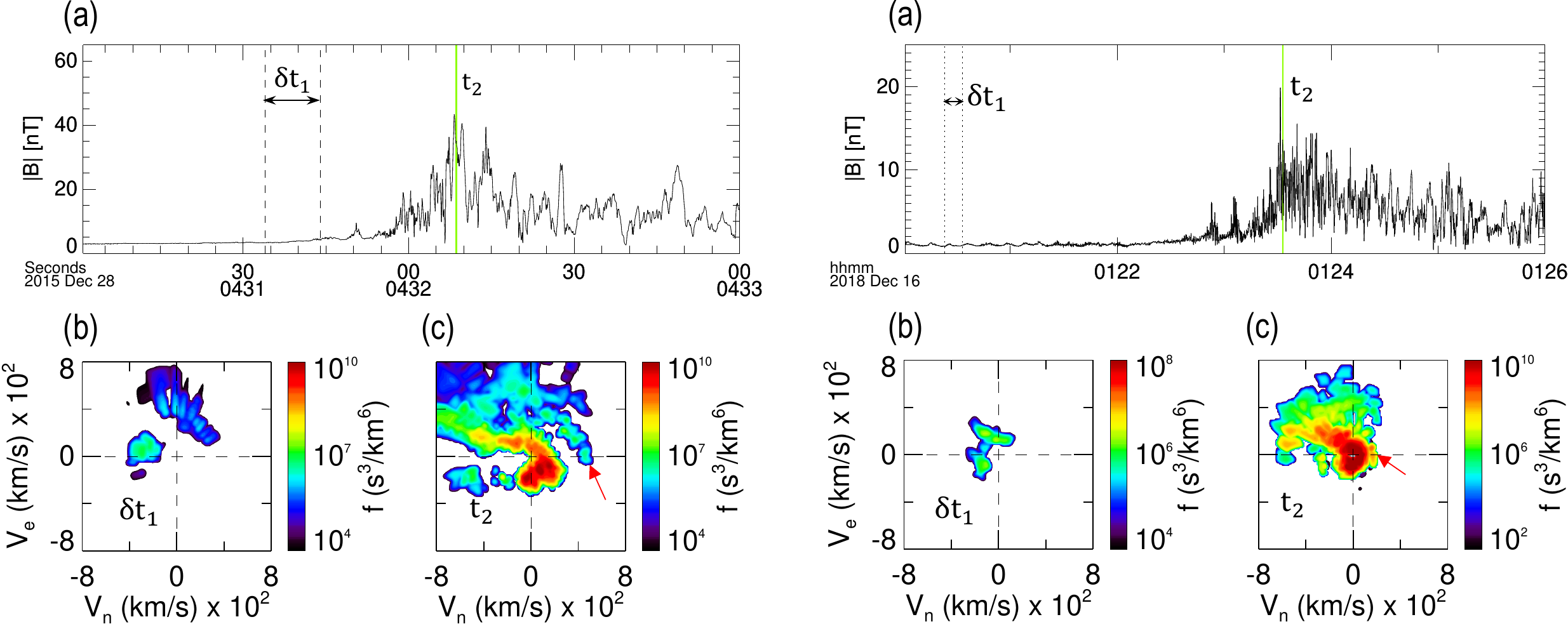}
\caption{Supporting figures for the shock events on 28 December 2015, 04:32:04 UT (Left), and on 16 December 2018, 01:23:32 UT (Right) based on MMS3 data. Panels show: (a) magnetic field profile across the shock, (b) HPCA He\textsuperscript{2+} distribution upstream of the shock during $\delta t_1$, and (c) FPI ion distribution at the shock front at time highlighted by $t_2$.
\label{fig:figSM2}}
\end{figure}
 
\clearpage

\bibliography{MendRefsMay.bib}

\end{document}